\documentclass[1p]{elsarticle}
\usepackage{float}
\usepackage{amsfonts}
\usepackage{amsmath}
\usepackage{amsthm}
\usepackage{amssymb}
\usepackage{commath}
\usepackage{cleveref}
\usepackage{comment}
\usepackage[hyphens]{url}
\usepackage{subfig}
\usepackage{tikz}
\usetikzlibrary{arrows.meta,datavisualization}
\biboptions{sort&compress}
\makeatletter
\def\ps@pprintTitle{%
 \let\@oddhead\@empty
 \let\@evenhead\@empty
 \def\@oddfoot{}%
 \let\@evenfoot\@oddfoot}
\makeatother

\begin{document}
%\maketitle

\begin{frontmatter}
\title{A short study comparing countries on the quality of response to the Covid-19 pandemic}

%% or include affiliations in footnotes:
%\begin{comment}
\author{Thilakam Venkatapathi}
%\ead{thilakammur@gmail.com}
\address{Department of Pathology, University of Kentucky, Lexington, KY 40536}

\author{Murugesan Venkatapathi*}
%\ead{murugesh@iisc.ac.in}
\address{Department  of  Computational  \&  Data Sciences, Indian Institute of Science, Bangalore 560012.}

\cortext[cor1]{Department  of  Computational  \&  Data Sciences, Indian Institute of Science, Bengaluru, India - 560012. Phone: +91 80 22933422. Email: murugesh@iisc.ac.in}

%\end{comment}
%\maketitle

\begin{abstract}
\textbf{Background}: We estimate the overall quality of response to the Covid-19 pandemic in the first 18 months, using a small number of known parameters and a proposed method that is reasonably robust to the uncertainties in the data.

\noindent \textbf{Methods}: The population-normalized values of deaths, diagnostic tests, confirmed cases, and doses of vaccines administered were considered. The average infection-fatality-rate provides us a baseline on potential deaths, and along with the test positivity rates in the formula, they add robustness to the estimates of the quality of response. 

\noindent \textbf{Results}: The scores are used to rank countries in two lists representing 84 large countries with a population greater than 10 million, and 85 countries with smaller populations. Additional possible corrections in the rankings of countries to include the per capita purchasing power and the age distribution, are also shown. In a supplementary note, an analysis of the robustness of the overall ranking list to the expected uncertainties in the data, and the maximum possible changes in the ranking of any country are presented.

\noindent \textbf{Conclusion}: In many countries, the outcomes are not significantly better than the baseline. A few significant inferences are pondered that may help unravel the causes of the poor outcomes.
\end{abstract}

\begin{keyword}
Covid-19; Countries; Quality of response.

\begin{comment}

\begin{itemize}
    \item \textit{What is already known on this topic} – The country-wise death count, positive cases and vaccination rates for Covid-19.
    \item \textit{What this study adds} – A comparative analysis using the population-normalized values of 6 different parameters.
    \item \textit{How this study might affect research, practice or policy} – Provides policy level insights on the global response to the Covid-19 pandemic and the country-wise outcomes.
\end{itemize}
\end{comment}
\end{keyword}

\end{frontmatter}
%\newpage

%\begin{comment}

\section{Introduction:}
With reported deaths due to the Covid-19 pandemic reaching 5 million it is the worst pandemic (sparing HIV/AIDS) since 1920, when a third of the world was infected and more than 20 million people lost their lives to the Spanish flu. When studying large complex problems that do not have a full description, in the interest of optimal solutions, we have a necessity to quantify the correlation of the outcomes with the inputs using a single number \cite{FairInequality2017}. It becomes further significant in the case of fighting pandemics where a small change in nature of the response by the governments and the people, can produce large changes in the outcomes. The unfortunate deaths due to the multiplicative nature of the communicable disease have an exponentially increasing or decreasing relation with time and the mitigating efforts. This motivate us for an evaluation of the average quality of the response to a pandemic as a single score, using a few measured parameters as highlighted in section 2. Next, the observations using the proposed relation and the publicly available data on the Covid-19 pandemic, are reported in section 3. This is followed by a discussion with notable points that may be explored further from a health policy and implementation point of view. A supplementary note presents the details of the scoring formula used, and a study to establish the robustness of the reported rankings in the presence of noisy data that simulates relatively large under-reported deaths or inefficient diagnostic tests, and this may be of interest to readers working on improving such evaluations.

\section{Methods:}
If we chose a formulation that minimises the scores for a higher quality of response, the scores assigned have to increase with the deaths reported due to the pandemic. The deaths considered are as fractions of the total population, deaths per million for example, to account for the varying populations of the countries. Direct interventions on the diseased such as effective treatments of the symptoms, or potential cures for the disease, indeed manifest as lower number of deaths. But the varying onsets and degrees of the pandemic in countries, and the mitigating response of the governments and people have to be considered. The deaths due to the disease alone do not qualify the response to the pandemic, when one has additional objectives such as protecting the livelihoods and the potential deaths due to other concomitant causes. For example, vaccination can prevent potential deaths in the future, and diagnostic testing furthers the effective treatments, and also helps in reducing the transmission of the disease if used along with the measures of social distancing. The number of tests, confirmed cases and vaccination doses used in the evaluation are also population-normalized values. Inclusion of tests and vaccinations in the formula for the quality of response also adds to its robustness when the deaths reported can have variations due to differing medical classifications and administrative efficiencies.  Particularly, this offsets the under-recording of deaths due to inadequate testing.

\begin{figure}[h]
		%\begin{center}
		\includegraphics[width = 15 cm]{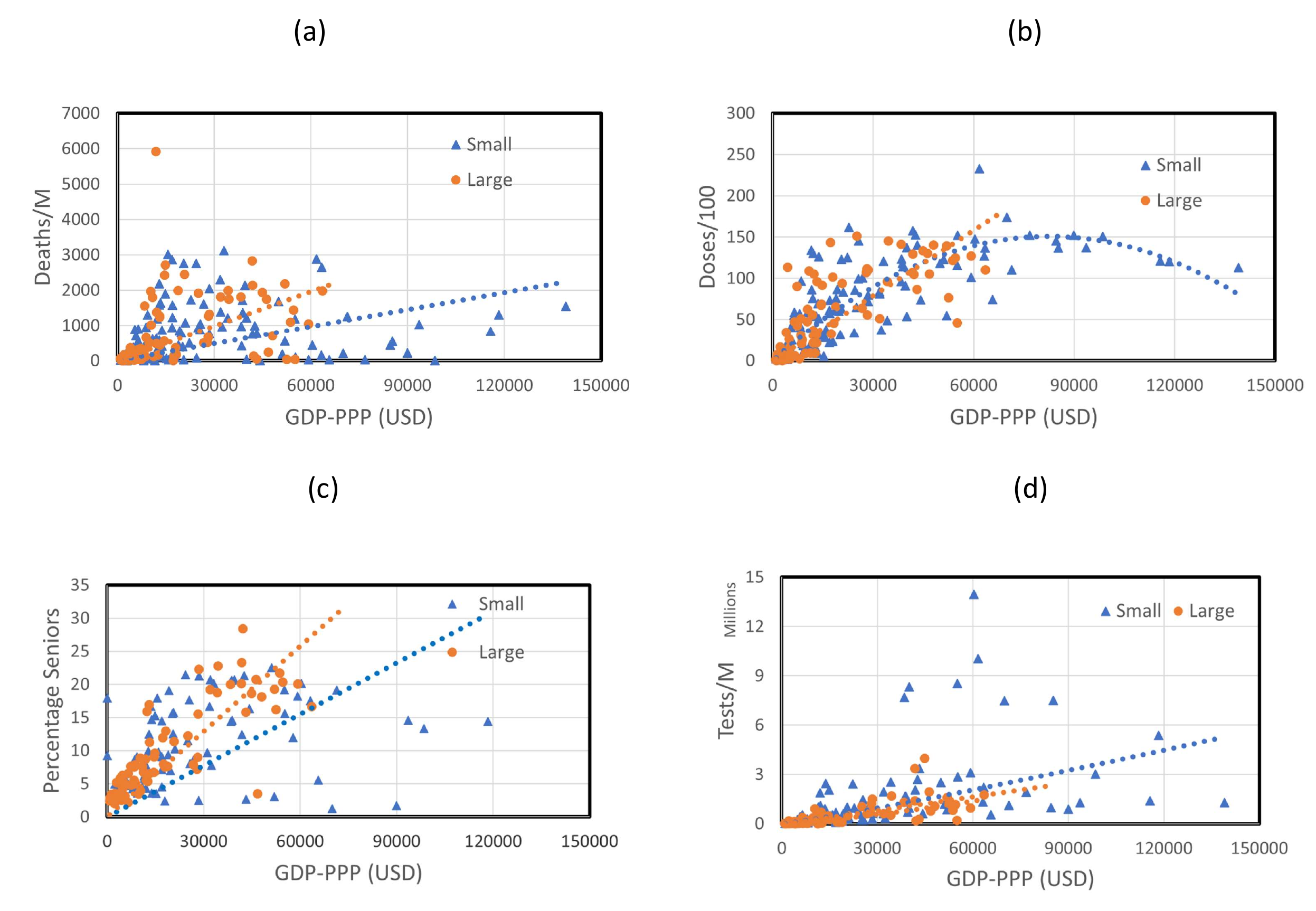}
			\caption{(a) Deaths per million; (b) Vaccine doses per 100 people; (c) Fraction of senior population in percentage; and (d) Tests per million;  All plotted with the corresponding per capita GDP values. Data sets `Small' and `Large' represents countries with populations less than or greater than 10 million people, respectively. The dashed lines are only guidelines to infer the correlations using a linear or polynomial fit for a least square error. }\label{fig:Correlations}
	    %\end{center}
	\end{figure}

Before we embark on building a formula, let us look at the correlation of the significant parameters with a known causal factor i.e. the correlation of deaths, diagnostic tests and vaccinations with the purchasing power in the population \cite{WInfo, WMeter, WBankData}. In this work we use the per capita GDP in purchasing power parity (PPP) terms reflecting the resources available to a person in the population, and not the per capita \textit{nominal} GDP estimated in dollar prices of goods, which is relatively less relevant. Counter-intuitively, the deaths have a strong positive correlation with per capita GDP of the country as shown in Figure 1a for both the small and large countries i.e. higher the per capita GDP, higher the likely number of deaths. Figure 1c unravels this causal relationship; higher the per capita GDP, higher is the life expectancy and the fraction of seniors ($>$ 65 yrs of age) in the population. It should be noted that the Covid-19 has infection-fatality-rates that increase exponentially with the age group (see figures 3 and 4 in \cite{ExpIFR}). The infection-fatality-rates represent the probability of death for an infected person, and it is different from the case-fatality-rates which considers only the confirmed cases. The former are typically estimated using the sero-prevalence of anti-bodies in the population, and less reliably using excess deaths observed, thus including the unreported infections as well. From a low infection-fatality-rate ($<$ $0.01\%$) for people less than 30 years of age, it increases exponentially to cross $1\%$ for ages above 65 \cite{IFR_multiple, ExpIFR}. The case-fatality-rates also set a reliable upper bound on the infection-fatality-rates. Figures 1a and 1c also show that the smaller countries (blue triangles) constitute two distinct groups - one with a higher GDP per capita and a correspondingly aged population, and another group of off-shore financial hubs, tourist destinations and oil-producing countries that have a high per capita GDP but a smaller fraction of senior population. Figures 1b and 1d also confirm the expected increase in testing and vaccination rates with the increase in the per capita GDP of a country. Deduction of a formula for scoring countries based on the correlation of these positive and negative factors with the purchasing power, is presented in the supplementary.

\section{Results:}
A score $S$ was used to estimate the quality of response to the Covid-19 pandemic, along with values reported until Aug 31, 2021 \cite{WInfo, WMeter}. While a delay of up to 4 weeks in reporting the number of the administered vaccination doses are possible, the other parameters were updated daily. The rankings of countries in tables \ref{fig:small_countries} and \ref{fig:large_countries} are based on $S$ with the lower scores representing a better quality of response and a higher ranking.

%\begin{comment}
\begin{equation} \label{eq:scoring_1}
    S = \log(1+FG)\text{ where }F=\frac{a_0D}{e^{-(\frac{a_0D}{D_0}+\frac{p}{p_0})}\sqrt{VT}} \text{ and }p=\frac{C}{T}
\end{equation}

where $G$ is the GDP per capita in PPP terms for the country scored, giving us the units of dollars per vaccine (or dollars per test) for the evaluated scores. 
Here $T$, $V$, $C$ and $D$ represent population-normalized values (per million people) of the tests, vaccine doses, confirmed cases, and deaths. $D_0$ represents the potential deaths when all the million people are infected in the absence of vaccines and the treatments available to the patients remain unaltered. For the \textit{actual} rankings of countries, any constant value of $G$ and $a_0$, like the average per capita values of the world and $a_0$=1 can be used. This implies that there is no compensation in scores for the varying economic conditions and ages of the population. Similarly, the maximum value of $p$ suggested by the World-Health-Organization (WHO) is $0.1$ beyond which tests become increasingly useless in breaking chains of transmission, and we use $0.1$ as the half life $p_0$ for tests. See the supplementary for the explanation and deduction of the above equation \eqref{eq:scoring_1} in estimating the scores.

For a compensated ranking with a hypothetical parity in economic and age factors of the populations, we consider the varying per capita GDP of the country, and the ratio of the fraction of the senior population globally ($>$ 65 years of age) and the fraction of seniors within the country \cite{WBankData}. Also, note that there is no reasonable way to compensate for the variations in geographies or the population densities of countries that impact the response to the pandemic. Thus, one may argue that the actual ranks are most appropriate for conclusions on the quality of response of the governments and the people, given the differing but known conditions they are subject to. The actual rankings and the compensated rankings have large differences only for some countries in West Asia where the per capita GDP is high but the fraction of senior population is low, and these countries lose significantly in rankings due to this compensation. Similarly, a few countries in Africa with a reasonable fraction of seniors but a low per capita GDP, gain significantly in this compensated ranking.

A study of the robustness of the ranking to the expected levels of noise in the data, is presented as a supplementary note. Under-recording of deaths and any inefficacy of reported tests can reduce the reliability of ranking, and reported values may vary up to $100\%$ even in well governed countries due to sharp waves of the pandemic when the medical and administrative systems get overwhelmed. For example, atleast a $25\%$ under-recording of Covid deaths was determined in the United States \cite{DeathCount1US, DeathCount2US}. But we note that indirect methods such as the estimation of a small fraction of total all-cause deaths as the excess unreported deaths due to the pandemic are not be reliable in general, as they are based on assumptions that may not be valid when we have a wide ranging impact on the economy and health systems. Note that the under-recording of deaths due to inadequate testing has already been addressed in the evaluated scores. Our study establishes that in the presence of additional \textit{up to} $100\%$ noise in the data, representing any other possible cause of under-reported deaths or inefficiency of tests, the average change in the ranking of the countries is less than 4. The average change in the rank is less than 5 for \textit{up to} $200\%$ additional noise in the data. These predictions include a confidence of $99\%$. Similarly, the maximum possible change in the rank of \textit{any} country in the list due to such noise is 19 and 28 respectively.
%\end{comment}

\setcounter{figure}{0}
\renewcommand{\figurename}{Table}
\begin{figure}[H]
%\begin{center}
    	\includegraphics[width = 14 cm]{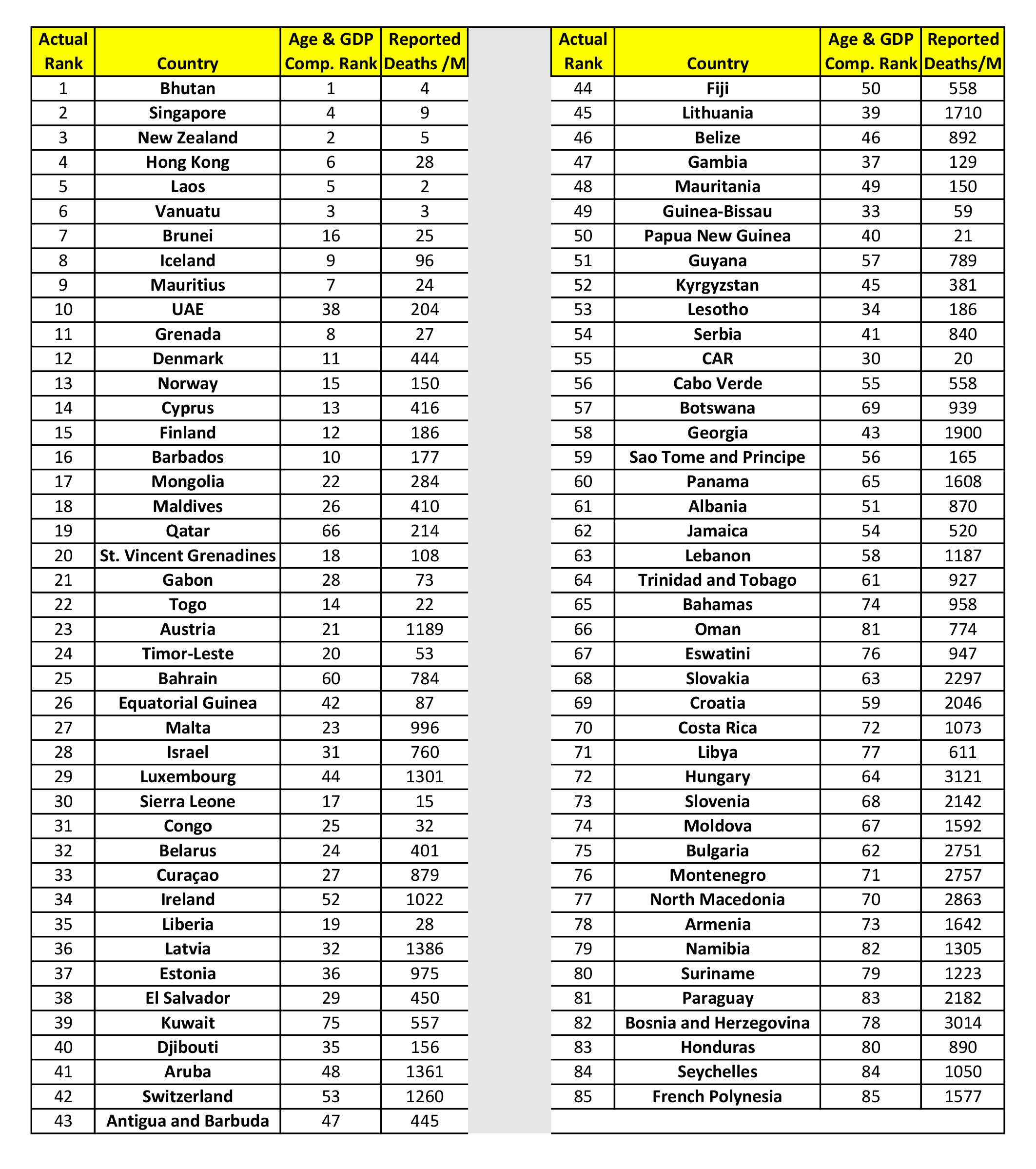}
		\caption{ List of small countries or administrative regions with a total population less than 10 million. Publicly available values \cite{WInfo, WMeter, WBankData} reported until Aug 31, 2021 were used in the evaluation, and the data is submitted along with the paper.}\label{fig:small_countries}
		%\end{center}
	\end{figure}

\begin{figure}[H]
    	\includegraphics[width = 14 cm]{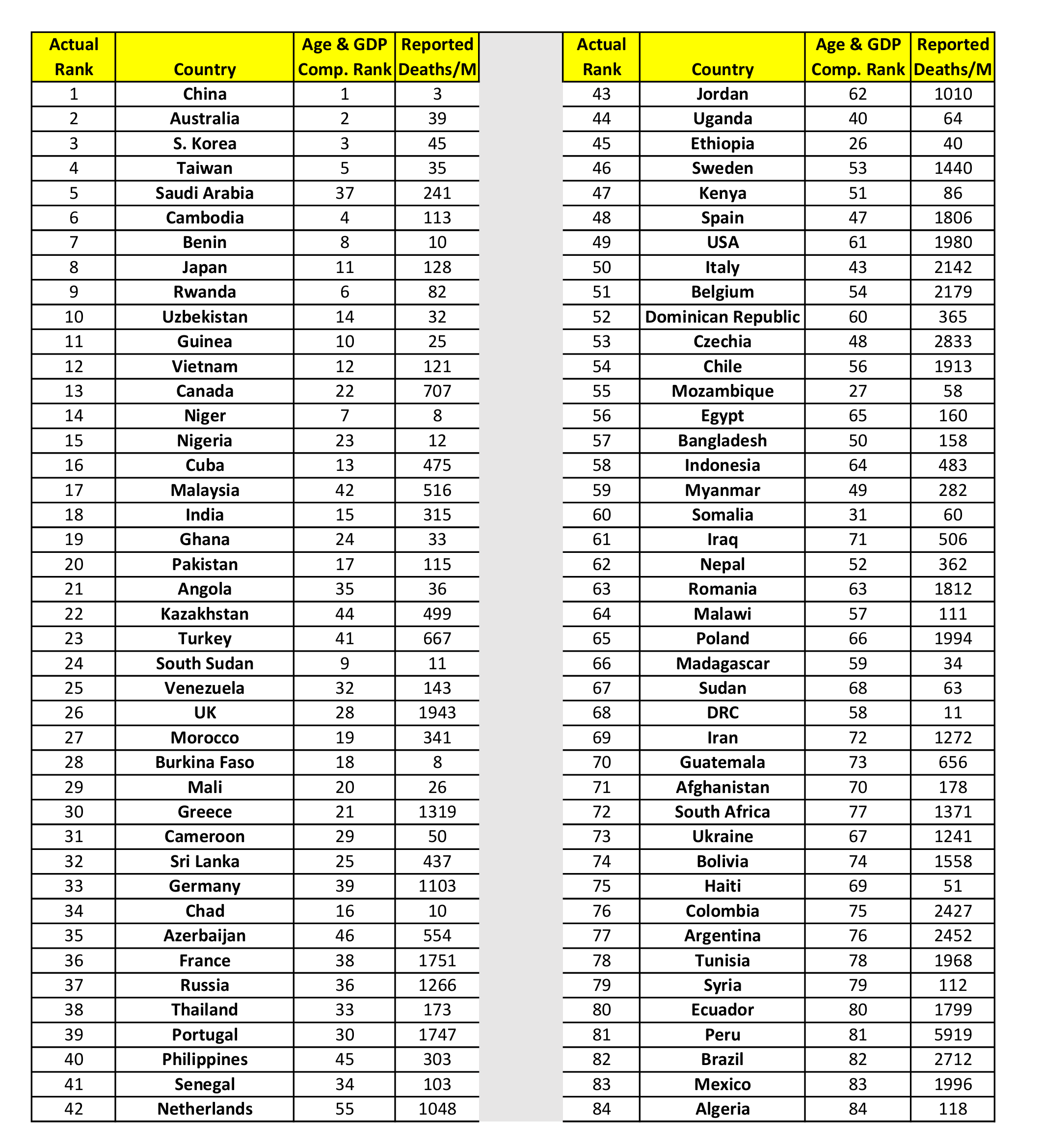}
		\caption{List of large countries with a total population greater than 10 million. Publicly available values \cite{WInfo, WMeter, WBankData} reported until Aug 31, 2021 were used in the evaluation, and the data is submitted along with the paper.}\label{fig:large_countries}
	\end{figure}

\section{Discussion:}
The ranking tables \ref{fig:small_countries} and \ref{fig:large_countries} of the small and large countries establish that the pandemic's devastation has touched all parts of the globe regardless of the economic, geographic and demographic variations. The top ranks have mostly been occupied by countries with low population densities and relatively insulated borders. It also shows that the mitigation efforts have not provided the results generally expected twelve months ago, given the no-response baseline and the known infection-mortality-rates determined for the unvaccinated people early-on during the pandemic. One has to wonder if the scientific and political efforts have been sub-optimal in ending the pandemic sooner. More specifically, well-designed administration of vaccine doses prioritized finely on age, and distributed based on population-density and the applied social distancing measures could have further reduced deaths, though it may have been politically unpopular. This might have as well released some of the evolutionary pressure on the virus to mutate into variants that can hide from our immune system, and further become dominant in the population \cite{Mutations2021}. Considering that the efficacy of the vaccines and the rates of vaccination were bound to be well below the required levels globally, scientific studies on optimal policies of vaccination based on these evolutionary aspects should have been emphasized. Note that when New York had the last outbreak of small pox in 1947, the entire population of 6.5 million had to be vaccinated within a month to get rid of the pandemic, and the Covid-19 virus has a similar rate of transmission. The expectation that people vaccinated under the Emergency-Use-Authorization would be tracked for quantitative assessments on the duration of immunity, has not been met as well. Easy to administer, more compliant and transmission-arresting nasal vaccines may provide a realistic longer-term option against the virus \cite{NasalV02020, NasalV12021}, that may otherwise require multiple intra-muscular vaccine doses for a person every year.

More surprising is the lack of effective anti-virals, repurposed drugs and other protocols for treatments of the symptoms and the disease itself. Questions on how one should establish `control' in randomized control trials in the times of a raging pandemic have risen, especially in the case of potential treatments where the meta-analyses of re-purposed therapeutics show clear benefits \cite{IverDec2020, IverMarch2021, IverApril2021, IverJune2021}. Note that in an ideal control trial, a notable fraction of the infected are supposed to be treated only with a placebo, which poses ethical dilemmas considering the risks associated in this pandemic \cite{Ethics1987Trials}. Such questions are many times intertwined with the economic impacts, and has led to inconsistencies both among the medical professionals and within the regulatory authorities, and resulted in ad hoc treatment protocols and advisories \cite{Ethics2020Trials, EfficacyTrials}. The unknown origin of the virus and the pandemic may also have had a huge effect on the scientific outcomes in finding a mitigation or cure for the disease \cite{OriginPandemic1}. Considering that the warming climate and intensive animal farming could be augmenting factors for future outbreaks of viral diseases \cite{Warming2005, WarmingVirusJapan, WarmingDiseaseResearch, Warming2020, WarmingNature}, this lack of understanding of the origin of the pandemic could be more costly in a future outbreak. There is always the possibility of a novel virus emerging that is as contagious as Covid-19 but with a higher rate of mortality. Recently, strong evidence of a potential laboratory role in the origin of Covid-19 has been reported \cite{Ambati2022Covid-origin}.

The second aspect that is not directly apparent in these evaluations but one that saved many lives, is the timely governance in the face of exponentially increasing cases in an acute wave of the pandemic. Bhutan stands out in its response though it may not be ranked highly in the human development index, and it should be no surprise for the readers familiar with its focus on social indicators such as \textit{Gross National Happiness} (GNH). Singapore with its strict measures of social distancing succeeds in keeping the deaths low despite a relatively high population density. Europe and South America are two continents that have been most impacted by the pandemic; with only few countries like Norway, Finland and Denmark being able to keep the pandemic relatively in check. The most populous country of China was expected to do well considering the permanent one-party rule and its unlimited power. The scientific prowess of China and its unique advantages as the reported origin of the pandemic may also have played a role in its excellent response to the pandemic and be ranked at the top. The intensity of the pandemic and the testing rates (a total of 65 confirmed cases and $\sim$110,000 tests per million) have been among the lowest in the world for more than a year. This has been attributed mostly to the high vaccination rates, but with the poor efficacy of its vaccines observed in Seychelles, parts of South America, and UAE, the data reported from China presents us with many contradictions. 

The other large country of India was expected to be consumed by the pandemic more than the smaller or the developed countries. The outcomes in India have been notably better than most parts of the world in its response to the pandemic, and ranked at 18 it is among the top quarter of countries. But, the devastating second wave of Covid-19 presented questions on the role of both the central/federal institutions on the issue of appropriate advisories and tracking of dangerous mutants \cite{DeltaVirus}, and the local governments at the states in not preparing for eventualities like the demands of oxygen. Note that cost and time effective solutions for generation of medical grade oxygen were described, available and mandated with appropriate licensing measures, a year before the onset of the second wave in India \cite{OxyWHO, OxyCDSCO, OxyMaha}. Unfortunately, the oxygen generating plants put into operation by the state/local governments after the second wave, may not be essential until a third wave of the pandemic \cite{OxyDelhi, OxyDelhiPlants}. A more appropriate response of the government and the people going into the second wave could have reduced its total death toll by a quarter, and India may have ranked higher at 14.

\section{Conclusion:}
The challenge of timeliness and \textit{quality} of response to a raging pandemic has put stringent demands on governments and people all around the world, and a delayed response of a great \textit{degree} has been shown to be of little consequence. The response forced on us has been equally challenging in the personal, economic and scientific spheres of activity. Significance of  timely monitoring of communicable diseases, transparent sharing of information among the global scientific community, along with the required scrutiny on the gain-of-function research on such pathogens, have been specifically highlighted. The contagious and mutating nature of the virus made national boundaries irrelevant, and it makes a strong case for a truly global response to such challenges in the future.

%\begin{comment}

\section{Declarations:}
\subsection*{Ethical approval and consent to participate:}
Not applicable. The study is based on anonymised population level meta-data freely available in public resources.

\subsection*{Consent for publication:}
Not Applicable.

\subsection*{Data Availability Statement:}
The used data is available as supplemental material along with the paper, for convenience of the reader. The publicly available sources of this data are also cited appropriately.

\subsection*{Competing interests:}
The authors declare that they have no conflicts of interest either financial or ethical in nature.

\subsection*{Funding:}
This research did not receive any specific grant from funding agencies in the public, commercial, or not-for-profit sectors.

\subsection*{Authors' contributions:}
All authors contributed to the data analysis and writing of the manuscript.

\subsection*{Acknowledgements:}
None.

%\end{comment}

%\printbibliography	
\bibliographystyle{unsrt}
\bibliography{reference.bib}

\begin{thebibliography}{10}

\bibitem{FairInequality2017}
Venkat Venkatasubramanian.
\newblock {\em How Much Inequality Is Fair?}
\newblock Columbia University Press, 2017.

\bibitem{WInfo}
Our world in data.
\newblock \url{https://ourworldindata.org/covid-vaccinations}.

\bibitem{WMeter}
Worldometer.
\newblock \url{https://www.worldometers.info}.

\bibitem{WBankData}
World {B}ank {O}pen {D}ata.
\newblock \url{https://data.worldbank.org}.

\bibitem{ExpIFR}
Andrew~T. Levin, William~P. Hanage, Nana Owusu-Boaitey, Kensington~B. Cochran,
  Seamus~P. Walsh, and Gideon Meyerowitz-Katz.
\newblock Assessing the age specificity of infection fatality rates for
  covid-19: systematic review, meta-analysis, and public policy implications.
\newblock {\em European Journal of Epidemiology}, 35(12):1123--1138, Dec 2020.

\bibitem{IFR_multiple}
John P.~A. Ioannidis.
\newblock Infection fatality rate of covid-19 inferred from seroprevalence
  data.
\newblock {\em Bull. World Health Organ.}, 99:19–33F, 2021.

\bibitem{DeathCount1US}
Daniel~M. Weinberger and et. al.
\newblock {Estimation of Excess Deaths Associated With the COVID-19 Pandemic in
  the United States, March to May 2020}.
\newblock {\em JAMA Internal Medicine}, 180(10):1336--1344, 2020.

\bibitem{DeathCount2US}
L.~M. Rossen, A.~M. Branum, F.~B. Ahmad, P.~D. Sutton, and R.~N. Anderson.
\newblock {Notes from the Field: Update on Excess Deaths Associated with the
  COVID-19 Pandemic — United States, January 26, 2020–February 27, 2021}.
\newblock {\em MMWR Morb. Mortal. Wkly. Rep. 2021}, 70:570--571, 2021.

\bibitem{Mutations2021}
Zijun Wang et~al.
\newblock {mRNA vaccine-elicited antibodies to SARS-CoV-2 and circulating
  variants}.
\newblock {\em Nature}, 592(7855):616--622, Apr 2021.

\bibitem{NasalV02020}
Ahmed~O. Hassan et~al.
\newblock A single-dose intranasal chad vaccine protects upper and lower
  respiratory tracts against sars-cov-2.
\newblock {\em Cell}, 183(1):169--184.e13, 2020.

\bibitem{NasalV12021}
Mattia Tiboni, Luca Casettari, and Lisbeth Illum.
\newblock Nasal vaccination against sars-cov-2: Synergistic or alternative to
  intramuscular vaccines?
\newblock {\em International Journal of Pharmaceutics}, 603:120686--120686, Jun
  2021.

\bibitem{IverDec2020}
S~Ahmed et~al.
\newblock A five-day course of ivermectin for the treatment of covid-19 may
  reduce the duration of illness.
\newblock {\em Int. J. Infect. Dis.}, 103:214--216, Dec 2020.

\bibitem{IverMarch2021}
H~Pott-Junior et~al.
\newblock Use of ivermectin in the treatment of covid-19: A pilot trial.
\newblock {\em Toxicol. Rep.}, 8:505--510, March 2021.

\bibitem{IverApril2021}
Pierre Kory, Gianfranco~Umberto Meduri, Joseph Varon, Jose Iglesias, and
  Paul~E. Marik.
\newblock Review of the emerging evidence demonstrating the efficacy of
  ivermectin in the prophylaxis and treatment of covid-19.
\newblock {\em American Journal of Therapeutics}, 28(3):e299--e318, Apr 2021.

\bibitem{IverJune2021}
Andrew Bryant, Theresa~A. Lawrie, Therese Dowswell, Edmund~J. Fordham, Scott
  Mitchell, Sarah~R. Hill, and Tony~C. Tham.
\newblock Ivermectin for prevention and treatment of covid-19 infection: A
  systematic review, meta-analysis, and trial sequential analysis to inform
  clinical guidelines.
\newblock {\em American Journal of Therapeutics}, 28(4):e434--e460, Jun 2021.

\bibitem{Ethics1987Trials}
Rosner F.
\newblock The ethics of randomized clinical trials.
\newblock {\em Am. J. Med.}, 82(2):283--290, 1987.

\bibitem{Ethics2020Trials}
Howard Bauchner and Phil~B. Fontanarosa.
\newblock {Randomized clinical trials and COVID-19: Managing Expectations}.
\newblock {\em JAMA}, 323(22):2262--2263, 2020.

\bibitem{EfficacyTrials}
V.~R. Emani et~al.
\newblock {Randomised controlled trials for COVID-19: evaluation of optimal
  randomisation methodologies-need for data validation of the completed trials
  and to improve ongoing and future randomised trial designs}.
\newblock {\em Int. J. Antimicrob. Agents}, 57(1):106222, 2021.

\bibitem{OriginPandemic1}
P.~Balaram.
\newblock Natural and unnatural history of the coronavirus: The uncertain path
  to the pandemic.
\newblock {\em Current Science}, 120(12):1820--1826, 2021.

\bibitem{Warming2005}
A.~A. Khasnis and M.~D. Nettleman.
\newblock Global warming and infectious disease.
\newblock {\em Arch. Med. Res.}, 36(6):689--696, Nov 2005.

\bibitem{WarmingVirusJapan}
Ichiro Kurane.
\newblock The effect of global warming on infectious diseases.
\newblock {\em Osong Public Health and Research Perspectives}, 1(1):4--9, Dec
  2010.

\bibitem{WarmingDiseaseResearch}
Lu~Liang and Peng Gong.
\newblock Climate change and human infectious diseases: A synthesis of research
  findings from global and spatio-temporal perspectives.
\newblock {\em Environment International}, 103:99--108, 2017.

\bibitem{Warming2020}
M.~B. Thomas.
\newblock Epidemics on the move: Climate change and infectious disease.
\newblock {\em PLoS Biol.}, 18(11):e3001013, 2020.

\bibitem{WarmingNature}
Xavier Rod{\'o}, Adri{\`a} San-Jos{\'e}, Karin Kirchgatter, and Leonardo
  L{\'o}pez.
\newblock Changing climate and the covid-19 pandemic: more than just heads or
  tails.
\newblock {\em Nature Medicine}, 27(4):576--579, Apr 2021.

\bibitem{Ambati2022Covid-origin}
Balamurali~K. Ambati, Akhil Varshney, Kenneth Lundstrom, Giorgio Palú,
  Bruce~D. Uhal, Vladimir~N. Uversky, and Adam~M. Brufsky.
\newblock Msh3 homology and potential recombination link to sars-cov-2 furin
  cleavage site.
\newblock {\em Frontiers in Virology}, 2, 2022.

\bibitem{DeltaVirus}
{Global Virus Network:}.
\newblock \url{https://gvn.org/covid-19/delta-b-1-617-2/}.

\bibitem{OxyWHO}
WHO~Interim Guidance:.
\newblock Oxygen sources and distribution for covid-19 treatment centres, April
  4-th, 2020.

\bibitem{OxyCDSCO}
CDSCO Notifications/Public-Notices:.
\newblock Granting permission to manufacturers of industrial oxygen to
  manufacture oxygen for medical use in the light of covid-19, April 7-th,
  2020.

\bibitem{OxyMaha}
Maharashtra {FDA} issues license to 7 more companies to produce medical oxygen,
  dated 9-th may 2020.
\newblock
  \url{https://thehealthmaster.com/2020/05/09/fda-issues-lic-to-7-more-cos-to-produce-medical-oxygen/}.

\bibitem{OxyDelhi}
Delhi Govt. Notification~No. DCI/PC/2021/321/1225:.
\newblock Medical oxygen production promotion policy of {D}elhi, August 19-th,
  2021.

\bibitem{OxyDelhiPlants}
Oxygen plants with 57mt capacity commissioned in {D}elhi, dated 9-th september
  2021.
\newblock
  \url{https://www.newindianexpress.com/cities/delhi/2021/sep/09/47-oxygen-plants-with-57-mt-capacity-commissioned-in-delhi-official-data-2356535.html}.

\end{thebibliography}
%\end{comment}
\end{document}